\documentclass[12pt,a4paper]{article}
\usepackage{amsmath,amssymb,url}
\usepackage{graphicx,tabularx}
\usepackage{array,url}
\usepackage{thophys}

\makeatletter
\def\preprintno#1{\def\@preprintno{#1}}
\def\address#1{\def\@address{#1}}
\def\email#1#2{\thanks{\tt #1@{}#2}}
\def\titlenote#1{\def\@titlenote{#1}}
\def\abstract#1{\def\@abstract{#1}}
\renewcommand\abstractname{ABSTRACT}

\newlength\preprintnoskip
\newlength\abstractwidth
\@titlepagetrue
\renewcommand\maketitle{\begin{titlepage}%
  \let\footnotesize\small
  \hfill\parbox{\preprintnoskip}{%
  \begin{flushright}\@preprintno\end{flushright}}\hspace*{1cm}
  \vskip 60\p@
  \begin{center}%
    {\Large\bf\boldmath \@title \par}\vskip 1cm%
    {\sc\@author \par}\vskip 3mm%
    {\@address \par}%
    \vskip 2cm
    {\small\@titlenote \par}%
  \end{center}\par
  \@thanks
  \vfill
  \begin{center}%
    \parbox{\abstractwidth}{\centerline{\abstractname}%
    \vskip 3mm%
    \@abstract}
  \end{center}
  \end{titlepage}%
  \setcounter{footnote}{0}%
  \let\thanks\relax\let\maketitle\relax
  \gdef\@thanks{}\gdef\@author{}\gdef\@address{}%
  \gdef\@title{}\gdef\@abstract{}\gdef\@preprintno{}
}%
\def\@citex[#1]#2{\if@filesw\immediate\write\@auxout{\string\citation{#2}}\fi
  \def\@citea{}\@cite{\@for\@citeb:=#2\do
    {\@citea\def\@citea{,\penalty\@m}\@ifundefined
       {b@\@citeb}{{\bf ?}\@warning
       {Citation `\@citeb' on page \thepage \space undefined}}%
\hbox{\csname b@\@citeb\endcsname}}}{#1}}
\def\citerange{\@ifnextchar [{\@tempswatrue\@citexr}{\@tempswafalse\@citexr[]}}
\def\@citexr[#1]#2{\if@filesw\immediate\write\@auxout{\string\citation{#2}}\fi
  \def\@citea{}\@cite{\@for\@citeb:=#2\do
    {\@citea\def\@citea{--\penalty\@m}\@ifundefined
       {b@\@citeb}{{\bf ?}\@warning
       {Citation `\@citeb' on page \thepage \space undefined}}%
\hbox{\csname b@\@citeb\endcsname}}}{#1}}
\long\def\@makecaption#1#2{%
  \sbox\@tempboxa{#1: \emph{#2}}%
  \ifdim \wd\@tempboxa >\hsize
    #1: \emph{#2}\par
  \else
    \hbox to\hsize{\hfil\box\@tempboxa\hfil}%
  \fi
  \vskip\belowcaptionskip}
\def\fmslash{\@ifnextchar[{\fmsl@sh}{\fmsl@sh[0mu]}}
\def\fmsl@sh[#1]#2{%
  \mathchoice
    {\@fmsl@sh\displaystyle{#1}{#2}}%
    {\@fmsl@sh\textstyle{#1}{#2}}%
    {\@fmsl@sh\scriptstyle{#1}{#2}}%
    {\@fmsl@sh\scriptscriptstyle{#1}{#2}}}
\def\@fmsl@sh#1#2#3{\m@th\ooalign{$\hfil#1\mkern#2/\hfil$\crcr$#1#3$}}
\makeatother

\newcommand\ltap{\
  \raise.3ex\hbox{$<$\kern-.75em\lower1ex\hbox{$\sim$}}\ }
\newcommand\gtap{\
  \raise.3ex\hbox{$>$\kern-.75em\lower1ex\hbox{$\sim$}}\ }

\newcommand\simge{\mathrel{%
   \rlap{\raise 0.511ex \hbox{$>$}}{\lower 0.511ex \hbox{$\sim$}}}}
\newcommand\simle{\mathrel{
   \rlap{\raise 0.511ex \hbox{$<$}}{\lower 0.511ex \hbox{$\sim$}}}}

\newcommand\be{\begin{equation}}
\newcommand\ee{\end{equation}}
\newcommand\bea{\begin{eqnarray}}
\newcommand\eea{\end{eqnarray}}
\newcommand\ba{\begin{array}}
\newcommand\ea{\end{array}}

\def\bq{\begin{equation}}
\def\eq{\end{equation}}
\def\ba{\begin{eqnarray}}
\def\ea{\end{eqnarray}}

\newcommand{\whizard}{\texttt{WHIZARD}}
\newcommand{\oMega}{\texttt{O'Mega}}
\newcommand{\ocaml}{\texttt{OCaml}}
\newcommand{\vamp}{\texttt{VAMP}}
\newcommand{\circe}{\texttt{CIRCE}}
\newcommand{\circeone}{\texttt{CIRCE1}}
\newcommand{\circetwo}{\texttt{CIRCE2}}
\newcommand{\feynrules}{\texttt{FeynRules}}
\newcommand{\guineapig}{\texttt{GuineaPig}}
\newcommand{\pythia}{\texttt{Pythia}}

\begin{document}

\date{\today}

\preprintno{DESY 14-045\\SI-HEP-2014-07}

\title{WHIZARD 2.2 for Linear Colliders}

\author{W.~Kilian\email{kilian}{physik.uni-siegen.de}$^a$, 
  F.~Bach\email{fabian.bach}{desy.de}$^b$,
  T.~Ohl\email{ohl}{physik.uni-wuerzburg.de}$^c$,
  J.~Reuter\email{juergen.reuter}{desy.de}$^b$}

\address{\it%
$^a$University of Siegen, Physics Department, \\
  Walter-Flex-Str. 3, D-57068 Siegen, Germany, 
\\[.5\baselineskip] 
$^b$DESY Theory Group, \\
  Notkestr. 85, D-22607 Hamburg, Germany
\\[.5\baselineskip]
$^c$University of W\"urzburg, Faculty of Physics and Astronomy, \\
  Am Hubland, D-97074 W\"urzburg, Germany}

\titlenote{%
  Talk presented 
  at the International Workshop on Future Linear Colliders (LCWS13),
  \\
  Tokyo, Japan, 11-15 November 2013
}

\abstract{
We review the current status of the \whizard\ event generator. 
We discuss, in particular, recent improvements and features that
are relevant for simulating the physics program at a future
Linear Collider. 
}

\maketitle


\section{Introduction}

\whizard~\cite{Kilian:2007gr} is a universal event generator for
elementary scattering and decay processes at high-energy colliders. 

\whizard\ contains the \oMega\ matrix element generator~\cite{omega}.
This program generates optimized code for multi-leg tree-level matrix
elements in the helicity formalism.  The multi-channel Monte-Carlo
integrator \vamp~\cite{Ohl:1998jn} takes care of high-dimensional
integrations.  The \vamp\ algorithm is adaptive both between and
within channels, and thus computes accurate phase-space integrals and
efficiently generates weighted and unweighted event samples.

The \whizard\ core connects these different components.  It contains
the algorithm for multi-channel phase-space parameterization and
mapping, it provides the user interface, interfaces to external
programs (e.g., parton distributions, event formats, hadronization),
the routines for writing and reading event files, and a parton-shower
module.  Furthermore, there are optional components for performing
numerical analysis and visualization of event samples.

For a realistic description of the ILC and CLIC environments,
\whizard\ contains the \circe~\cite{Ohl:1996fi}
package for beam-spectrum simulation.  Alternatively, it can
digest \guineapig\ beam-event samples.

\section{Development}

\subsubsection*{\whizard~1}

\whizard\ was initiated in the context of the TESLA design
study~\cite{Kilian:2001qz}.  It started in the 1990s as a project for
an improved calculation of electroweak processes\footnote{namely, W,
  Higgs, Z and Respective Decays, thus the acronym} at high-energy
lepton colliders, where full multi-leg matrix elements without
factorization were required.  During the following decade, it was
developed towards a universal generator for partonic events at lepton
and hadron colliders.

Between 2000 and 2010, important improvements included the implementation
of QCD color and thus the full Standard Model (SM), parton
distributions, event samples, and support for a growing list of models
beyond the SM (BSM).  Among the latter, most prominent is the
implementation of the full MSSM and further variants of SUSY models
such as the NMSSM~\cite{Reuter:2009ex}.  \whizard\ supports the SUSY Les
Houches Accord (SLHA)~\cite{SLHA} for interfacing SUSY spectrum
generators.  The package has been used for many theory and experimental
studies for LHC and TESLA/ILC (see e.g.~\cite{WHIZARD_SUSY}).  In
particular, it was used for generating the Linear Collider event
database at SLAC.

\whizard~1 is no longer actively supported.  Development stopped with
version 1.94 in 2010.  The follow-up releases until 1.97
include only bug and regression fixes and the finalization of the
manual.

\subsubsection*{\whizard~2}

A thorough rewrite of the core program became necessary for meeting
the increased demands regarding applications and versatility of
\whizard\ as a tool for phenomenology and experiment.  This resulted
in the \whizard~2 series of releases.

In \whizard~2, the static setup of input files has been replaced by a
domain-specific language SINDARIN.  This user interface allows for
simple and compact input in standard use cases, but it also equips the
user with the full power of a programming language, adapted to the
special needs and conventions of particle physics.  It deals with
topics such as cuts, analysis, interfaces, process collections, or
parameter scans.

\whizard~2 make use of OpenMP for parallel execution.  New models can
be added using the \feynrules~\cite{Christensen:2008py}
package~\cite{Christensen:2010wz}.  Internally, calculations are
performed using a generic density-matrix formalism.  Furthermore, the
program contains a parton-shower module~\cite{Kilian:2011ka} matched
with exact matrix elements via the MLM scheme, as an alternative to
external showering.

\subsubsection*{The Package}

The programming language of \whizard~2 is modern Fortran (Fortran
2003), as it is supported by current compilers including the free
\texttt{gfortran}, version 4.7 and higher.  (The subpackage \oMega\ is
written in the functional programming language \ocaml, which
is freely available for all major platforms.)  The current \whizard\
production version is 2.1.1 (as of 2012).  The complete package works
and can be installed on all recent Linux and Mac OS systems.  The new
version 2.2.0 is currently (March 2014) in beta stage and due to
replace the previous version.


\section{Technical Improvements}
\label{sec:techphys}

\subsubsection*{Workflow}

The workflow in \whizard~2 has been greatly simplified compared to
\whizard~1, which did involve a conglomerate of shell and Perl
scripts, Makefiles and staged compilation.

The installation is possible at a central place, or locally in a user
directory.  It conforms to the standard toolchain of
\texttt{automake}, \texttt{autoconf}, and \texttt{libtool}.  The
program is maintained at the HepForge server~\cite{manual} and can be
downloaded either as a tagged \texttt{.tar.gz} version, or as a
development version from the public \texttt{svn} repository.  Installation
of the whole package proceeds via the usual \texttt{configure} --
\texttt{make} -- \texttt{make install} triple and the optional
\texttt{make check} and \texttt{make installcheck} steps for
additional peace of mind.

User projects can then be set up in arbitrary directories without any
predefined structure.  Usually, there is a single command
\texttt{whizard} and a single SINDARIN input file.  Alternative modes
of using \whizard\ include an interactive mode, or linking it as a
subroutine library that is interoperable with C, C++ or any other
C-compatible language (e.g., Python).  \whizard\ generates and digests
process code on-the-fly in form of dynamically linked libraries.  For
working with batch clusters, a statically linked mode is also
available.

\subsubsection*{Internal Structure}

\whizard~2 is broken down into modules according to the modern Fortran
standard.  Except for some legacy parts, the program is written in a
strict object-oriented fashion.  Between versions 2.1 and 2.2, there
has been a further major refactoring of the code which introduced an
extra abstraction layer for the central types and objects.  There is
now a consequent separation of interface from implementation, which
greatly improves the maintainability and, in particular, the
possibilities for future module replacements, reimplementations, and
extensions.  This affects, for instance, process structure, matrix
element calculation methods, beam structure, integration methods,
decays and shower.

\subsubsection*{Maintenance}

The \whizard\ system is maintained with \texttt{svn} version control
at HepForge.  All commits are run automatically through a chain of
several hundreds of unit and function tests, using a \texttt{Jenkins}
continuous integration server.  Bugs and feature requests are handled
using HepForge's tracking system.


\section{Physics}

\subsubsection*{\oMega}

The matrix element generator \oMega~\cite{omega} is able to
generate exact tree-level matrix elements with multiple external
legs (successful tests on standard hardware have involved up to 15 external
particles).  It is based on a recursive algorithm that reuses
\emph{all} common subexpressions and replaces the forest of all tree
diagrams by the equivalent directed acyclical graph (DAG).
One of the major improvements in
version 2 is the treatment of color for QCD amplitudes.  It uses the
color-flow formalism with phantom $U(1)$
particles~\cite{Kilian:2012pz} as an efficient way to generate
colorized DAGs that can be evaluated exactly or be projected onto
color-flow amplitudes.

\subsubsection*{Process Definition}

\whizard\ now implements the possibility to let processes consist of
several different components, which enables one to define process
containers for inclusive production samples.  This is technically
different from flavor sums (also available), where masses have to be
identical in order to use the same phase space.  In \whizard\ 2.2,
this feature will be made available to the user by appropriate syntax
in SINDARIN, covering inclusive processes and a detailed
specification of decay chains.

\subsubsection*{Process Evaluation}

The internal density-matrix formalism keeps track of
exact spin and color correlations.

Apart from complete matrix elements, a new feature of \whizard~2 is
the possibility to generate decay chains and cascades.  They are
composed of an arbitrary choice of elementary processes which are
separately integrated, but concatenated for the event generation.  For
the intermediate states, the default is to take full spin correlations
into account.  There is also an option to restrict to classical spin
correlations (only diagonal entries in the spin-density matrix), or to
even switch off spin correlations completely.  If requested, \whizard\
will set up decays and branching fractions for the
chosen model automatically.

\whizard~2 allows for arbitrary factorization and renormalization
scale settings (affecting QCD); the syntax parallels the one for cuts
or analysis.  

\subsubsection*{Reweighting}

Another new feature within \whizard\ 2.2 is the possibility to
reweight existing event samples, generated either internally or read
from file, when changing the setup of the original process,
e.g. parameters in the hard matrix elements, the event scale, the
chosen structure functions, or the QCD parton shower.


\section{Linear Collider Simulation}
\label{sec:lcsim}

\whizard\ has been developed as a generator for any high-energy
collider, but it contains modules that are dedicated to the
description of a lepton collider environment.  In particular,
given the required level of precision at ILC and CLIC, an accurate
description of beam
properties is essential for studies and analyses.

\subsubsection*{Beamstrahlung}

The program uses an interface to the \circeone~\cite{Ohl:1996fi}
program that parameterizes the beams of an $e^+e^-$ collider and
provides an event generator for factorized beam spectra.  Interfacing
this generator (or, alternatively, the parameterized spectrum
directly), \whizard\ can integrate and simulate any $e^+e^-$ process
with a realistic beam description.  As an upgrade to previous
versions, \whizard~2.2 ships with beam spectra that correspond to the
current ILC design parameters. To account for cases where such a
factorized form is insufficient, \whizard\ can alternatively read
beam-event files as they are produced by \guineapig.

\subsubsection*{ISR}

On top of the beamstrahlung effects, lepton-collider processes are
strongly affected by electromagnetic initial-state radiation (ISR).
\whizard\ implements ISR in a standard structure-function formalism
that resums the corrections from infrared (leading) and collinear (3rd
order) radiation and implements them in kinematics and dynamics, if
requested.

\subsubsection*{Polarization}

Furthermore, \whizard\ allows for specifying beam polarization,
ranging from unpolarized, left- or right-handed circular or
transversal polarization to arbitrary spin-density matrices.  The
polarization and polarization fractions are specified for both beams
independently.  The user can also define asymmetric beam setups and a
crossing angle, which will be taken into account in the kinematics
setup.

\subsubsection*{Photons}

Photons as initial particles are available in various incarnations:
on-shell, radiated from $e^\pm$ (effective photon approximation), or
beamstrahlung photons generated by \circeone.  A photon-collider
option that uses the \circetwo\ beam description is also available but
no longer maintained, due to the lack of current ILC or CLIC
photon-collider mode beam parameters and simulations of the
corresponding beam-beam interactions.

\subsubsection*{Events}

For cuts, reweighting and internal analysis, \whizard\ employs its
dedicated language SINDARIN that allows for computing a wide range of
event-specific and generic observables.

\whizard\ supports various event output formats, including the
traditional HEPEVT format and its derivatives, StdHEP, LHEF, HepMC,
and others.  A direct interface to LCIO is planned.


\section{QCD}
\label{sec:qcd}

\subsubsection*{Parton Shower}

For a precise calculation for exclusive Linear Collider processes, a
program needs fine control over QCD corrections.  Regarding real
radiation, the multi-leg capability of \whizard\ allows for including
high orders of the QCD coupling.  Collinear and soft radiation in
exclusive events is affected by large
logarithms, which are conveniently resummed in the semi-classical
approach of a parton shower algorithm.

Using standard event formats and suitable cuts, \whizard\ allows for
attaching an external parton-shower generator.  \whizard~2
furthermore contains an internal showering module in two different
incarnations: a $k_T$ ordered shower along the lines of the \pythia\
shower~\cite{Sjostrand:2006za}, and an analytic parton shower, which
keeps the complete shower history and allows to reweight
it~\cite{Kilian:2011ka}.  There is support for combining exact matrix
elements and QCD radiation from the parton shower using the MLM
matching prescription.  These modules are forseen to receive a more
detailed validation, tuning and further improvements after the 2.2
release.

Beyond the (partonic) parton shower, hadronization and hadronic decays
are not performed by \whizard\ internally, but can be applied to the
generated partonic event samples via the \pythia~6
package~\cite{Sjostrand:2006za} which is attached to the \whizard\
distribution, or using e.g. LHE event samples that are then fed into a
different external (shower and) hadronization package.

\subsubsection*{Virtual Corrections and Subtraction}

\whizard~2 with \oMega\ matrix elements is an event generator of
tree-level processes.  There have been several projects that extended
it to next-to-leading order, including loop corrections and proper
infrared-collinear subtraction.
Ref.~\cite{Robens:2008sa,Kilian:2006cj} describes the extension of
\whizard~1 to a positive-definite NLO event generator for the
electroweak pair production of charginos in the MSSM, including full
electroweak SUSY corrections matched to the photon initial and final
state radiation.  Independently,
Ref.~\cite{Binoth:2009rv,Greiner:2011mp} implemented the QCD NLO
correction with subtraction for a particular LHC process.  Along these
lines, the Binoth Les Houches Accord (BLHA) interface~\cite{BLHA} has
been implemented for reading and writing contract files with one-loop
programs (OLP), and has been validated.

Building upon the new data structures of \whizard~2.2, an
implementation of automatic NLO QCD corrections is currently being
developed.

\subsubsection*{Top-Quark Threshold}

\begin{figure*}
 \includegraphics[scale=0.63]{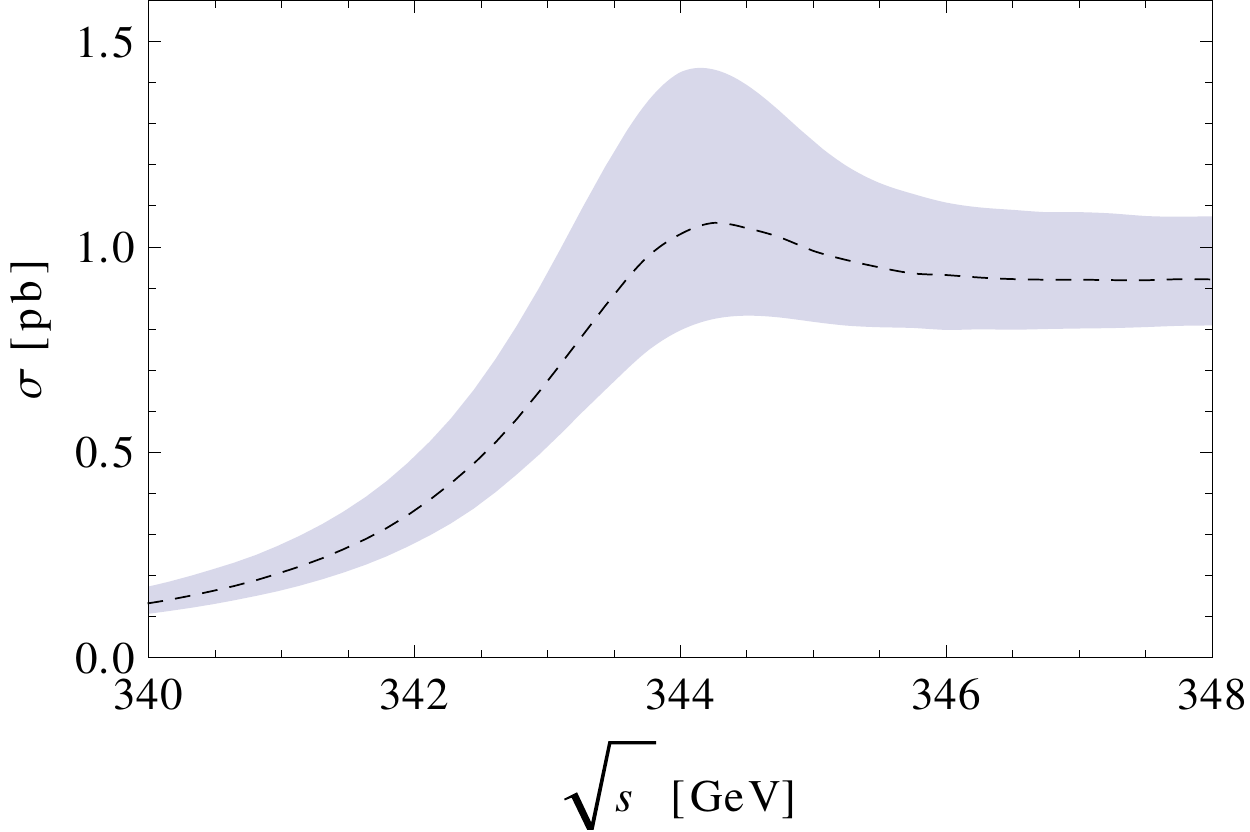}
 \hfill{}
 \includegraphics[scale=0.48]{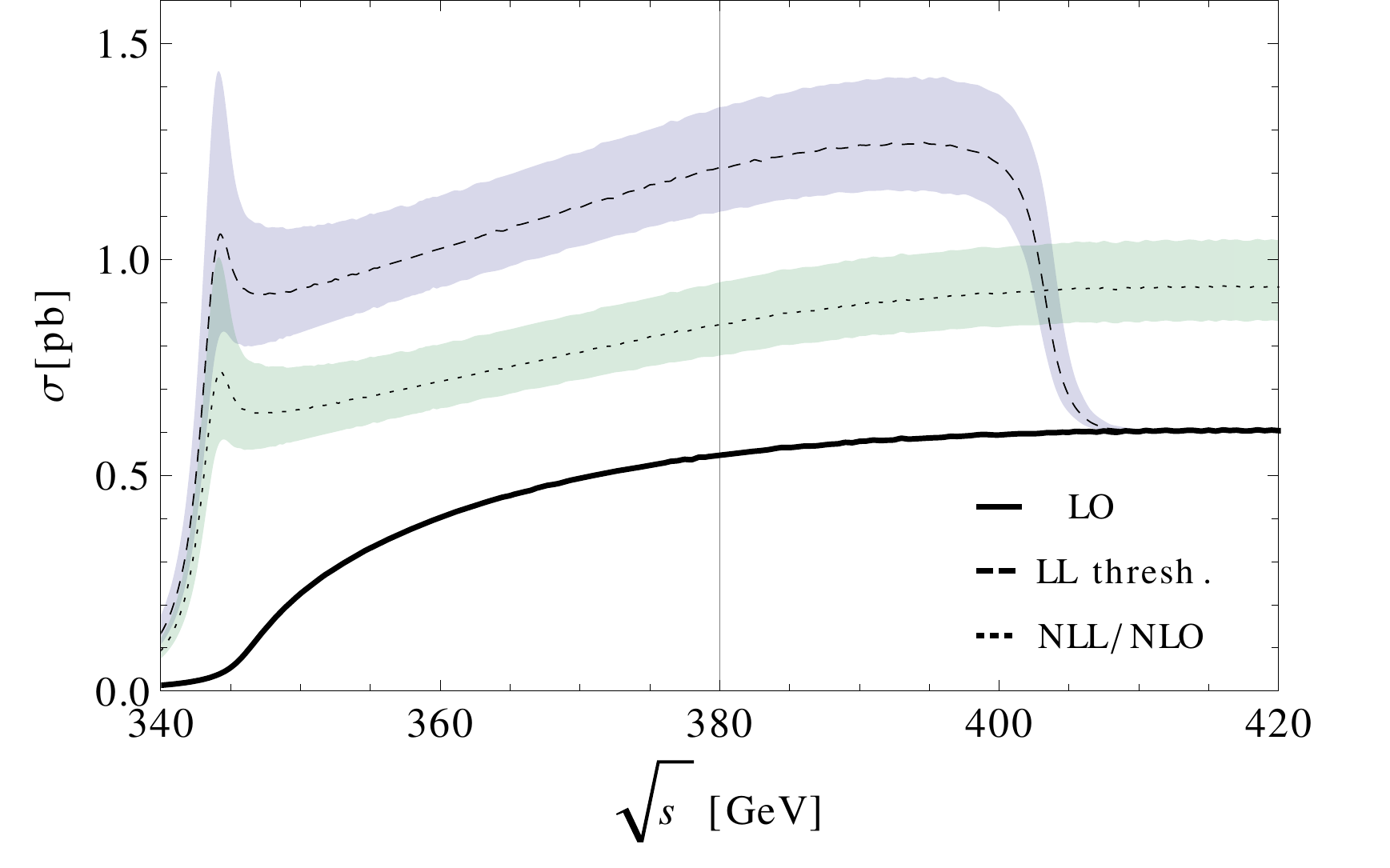}
 \caption{\whizard\ total $e^+e^-\to t\bar t\to b\bar bW^+W^-$ cross
 section including non-relativistic threshold corrections at LL order:
 threshold region (left) reproducing the shape e.~g.~in~\cite{Hoang:2013uda},
 and matching to continuum (right),
 which is at the current status of progress achieved by adapting K-factors
 for NLL threshold and NLO continuum normalization
 (uncertainties from variation of the soft renormalization scale).
 \label{tt_thresh}}
\end{figure*}

A high-luminosity linear collider will be capable of a high-precision
scan of the top-quark pair-production threshold~\cite{ILCTDR}.  To
match this on the theoretical side, one needs to resum logarithms of
the top velocity $\sim\alpha_s\ln v$ as well as gluon Coulomb
potential terms $\sim\alpha_s/v$ in a non-relativistic approach and to
relate this to the relativistic matrix elements in the continuum.
There is an ongoing project for including these effects in \whizard~2
which will make the theoretical calculation available in the
simulation of exclusive events.  As a first step, the
leading-logarithmic approximation matched to leading-order matrix
elements using a relative K-factor has already been implemented,
cf.~Fig.~\ref{tt_thresh}, and will be included in an upcoming
release~\cite{topthreshold}.

\section{Physics models}
\label{sec:physics}

\subsubsection*{Particles}

As a generator of hard matrix elements, \whizard\ has to support
various particle species and interactions.  The allowed spin
representations for particles are 0, 1, 2 (bosons) and 1/2, 3/2
(fermions), all massive or massless, both Dirac and Majorana spinors,
optionally colored (triplet or octet).  The \whizard\ libraries
support all Lorentz structures for interactions in the models
described below.  A a completely general framework supporting all
possible Lorentz structures is under construction.

\subsubsection*{BSM Models}

Beyond the SM and its QCD and QED subsets, \whizard\ supports the
minimal supersymmetric Standard Model (MSSM) with different variants
and extensions.  These include models with
gravitinos~\cite{WHIZARD_SUSY} and the
NMSSM~\cite{Reuter:2009ex} (see also~\cite{Reuter:2010nx}). 

Among models with strongly interacting sectors \whizard\ includes
Little Higgs models in different incarnations, with and without
discrete symmetries, cf.~\cite{WHIZARD_LHM}.

\whizard\ has also been used for studying more exotic models such as
the noncommutative SM~\cite{WHIZARD_NC} (not included in the official
release).  It further supports the completely general two-Higgs
doublet model (2HDM), as well as generic models containing a $Z'$
state, and extra-dimensional models like Universal Extra Dimensions
(UED). A more detailed list can be found in the \whizard\
documentation~\cite{manual}.

\subsubsection*{Effective Theories}

As an alternative tool for studying deviations from the SM, \whizard\
contains SM extensions with anomalous couplings, expressible as
coefficients of higher-dimensional operators in an effective theory.
Several models in \whizard's library define either anomalous triple
and quartic gauge boson couplings, which have been used for studies at
LCs~\cite{Beyer:2006hx} or at
LHC~\cite{Boos:1997gw,Alboteanu:2008my,Reuter:2013gla}.  Anomalous
top-quark couplings are also supported~\cite{Bach:2012fb}.

Recent interest in the physics of high-energy vector-boson scattering
has triggered the development and addition of simplified models for
strong interactions and compositeness (SSC)~\cite{kmatrix}.  In
addition to generic new degrees of freedom, they implement a
unitarization procedure that is required for extrapolating into the
energy range that will become accessible at ILC and, in particular, at
CLIC.


\section{Conclusion and Outlook}
\label{sec:outlook}

\whizard\ is a versatile and user-friendly tool for both SM and BSM
physics at all kinds of high-energy colliders.  Special emphasis has
always been put on the particular requirements for an accurate
simulation of exclusive events in the linear-collider environment.
This involves a detailed account of the nontrivial beam properties,
and of polarized and non-collinear beam configurations.

The \whizard\ collaboration actively pursues the further development
of the package, in particular regarding the physics program at ILC and
CLIC.  This required a thorough refactoring for version 2.2, which
subsequently will be the base for further development.  Plans for new
features include the support for more general Lorentz and color
structures in models, a more convenient model interface,
power-counting of coupling constants in the matrix element, further
refinements in the beam description, and automatic support for
higher-order corrections.


\section*{Acknowledgments}

JRR has been partially supported by the Strategic Alliance for
Terascale Physics of the Helmholtz-Gemeinschaft.
TO is supported by the German Ministry of Education and Research
(BMBF). WK and JRR want to
thank the organizers for scheduling the conference in the amazing
Japanese autumn foliage season.


\end{document}